\documentclass[12pt,english,showpacs,superscriptaddress]{revtex4}

\usepackage{amsmath}
\usepackage{amssymb}
\usepackage{esint}
\usepackage{babel}



\begin{document}

\title{Dynamics of a Dirac Fermion in the presence of spin noncommutativity}

\author{A. F. Ferrari}
\email{alysson.ferrari@ufabc.edu.br}
\affiliation{Universidade Federal do ABC, Centro de Ci\^encias Naturais e Humanas,
Rua Santa Ad\'elia, 166, 09210-170, Santo Andr\'e, SP, Brasil}

\author{M. Gomes}
\email{mgomes@fma.if.usp.br}
\affiliation{Instituto de F\'\i sica, Universidade de S\~ao Paulo, Caixa Postal
66318, 05315-970, S\~ao Paulo - SP, Brazil}

\author{V. G. Kupriyanov}
\email{vladislav.kupriyanov@gmail.com}
\affiliation{Universidade Federal do ABC, Centro de Matem\'atica, Computa\c{c}\~ao
e Cogni\c{c}\~ao, Rua Santa Ad\'elia, 166, 09210-170, Santo Andr\'e,
SP, Brasil}

\author{C. A. Stechhahn}
\email{carlos@fma.if.usp.br}
\affiliation{Instituto de F\'\i sica, Universidade de S\~ao Paulo, Caixa Postal
66318, 05315-970, S\~ao Paulo - SP, Brazil}


\begin{abstract}
Recently, it has been proposed a spacetime noncommutativity that involves
spin degrees of freedom, here called ``spin noncommutativity''.
One of the motivations for such a construction is that it preserves
Lorentz invariance, which is deformed or simply broken in other approaches
to spacetime noncommutativity. In this work, we gain further insight
in the physical aspects of the spin noncommutativity. The noncommutative
Dirac equation is derived from an action principle, and it is found
to lead to the conservation of a modified current, which involves
the background electromagnetic field. Finally, we study the Landau
problem in the presence of spin noncommutativity. For this scenario
of a constant magnetic field, we are able to derive a simple Hermitean
non-commutative correction to the Hamiltonian operator, and show that
the degeneracy of the excited states is lifted by the noncommutativity
at the second order or perturbation theory.
\end{abstract}

\maketitle

\section{Introduction}

The idea that spacetime may be noncommutative at very small scales
has its roots in semiclassical arguments stating that the principles
of quantum mechanics and general relativity together imply in an absolute
limit in the localization of events near the Planck scale\,\cite{doplicher:1994tu}.
The various instances of noncommutative spaces that have been studied
in the literature in the last decades, therefore, represent an effort
in uncovering some of the properties of spacetime at very small length
scales, so gaining some understanding about quantum gravity. One usually
expects physical effects related to quantum gravity to appear only
in very high-energy processes, where quantum field theory is the most
adequate theoretical tool. However, the study of relativistic or even
nonrelativistic quantum mechanics with noncommutative coordinates
has the advantage of exploring the noncommutativity of coordinates
in a simpler setting, better clarifying its physical consequences. 

In this context, various possibilities may arise, the simpler one
being when the noncommutativity is parametrized by some constant matrix.
This so-called ``canonical noncommutativity'' became quite popular
since the discovery of its connection with string theory\,\cite{seiberg:1999vs}.
Quantum mechanics with canonical noncommutativity is defined by the
commutation relations
\begin{equation}
\left[\hat{x}_{i}\,,\,\hat{x}_{j}\right]=i\theta_{ij}\,,\label{eq:canonicalNC}
\end{equation}
and it is implemented by means of the change of variables
\begin{equation}
\hat{x}_{i}=x_{i}-\frac{1}{2}\sum_{j}\theta_{ij}p_{j}\,\,,\,\,\hat{p}_{i}=p_{i}\,,
\end{equation}
where $x_{i}$ and $p_{i}$ are operators satisfying the standard
commutation relations of quantum mechanics, and $i,j=1,2,3$. In this
way, the Schroedinger equation in noncommutative space has the standard
form, but involves the modified potential
\begin{equation}
V\left(x_{i}-\frac{1}{2}\sum_{j}\theta_{ij}p_{j}\right)\,.
\end{equation}
Specific quantum mechanical potentials may then be studied using standard
perturbation theory\,\cite{gamboa:2000yq,chaichian:2000si,nair:2000ii}
or $1/N$ expansion\,\cite{Ferrari:2010en}, for example. One shortcoming
of this approach is that Lorentz -- or rotational, in the non-relativistic
case -- symmetry is generally lost since the constant $\theta_{ij}$
may define a preferred direction in space. For other aspects of noncommutative
quantum mechanics, see for example\,\cite{Adorno:2011wj,Gitman:2007ip,Vassilevich:2006tc,Deriglazov:2002gi,Deriglazov:2002xc,Deriglazov:2002zz,ferrari:2007vc}.

One may find in the literature several alternative approaches which
does not suffer from this symmetry loss -- such as, for example, Snyder's
work of 1947\,\cite{snyder:1946qz}, usually referred to as the first
proposal of a noncommutative spacetime. There, the commutator of two
coordinates is proportional to the Lorentz generator, 
\begin{equation}
\left[\hat{x}_{\mu}\,,\,\hat{x}_{\nu}\right]=\frac{i}{\Lambda^{2}}M_{\mu\nu}\,,\label{eq:snydersNC}
\end{equation}
where $\Lambda$ is some large UV scale. The Synder's algebra preserves
Lorentz invariance as it involves only covariant objects. For some
recents developments regarding Snyder's noncommutativity, see
for example\,\cite{Mignemi:2011gr,Lu:2011it,Lu:2011fh,Banerjee:2006wf}.
Closer to the canonical noncommutativity proposal is the idea that
relation (\ref{eq:canonicalNC}) is compatible with a twisted Lorentz
symmetry, understood as a Hopf algebraic symmetry with a non-trivial
coproduct~\,\cite{chaichian:2004za,balachandran:2005eb,balachandran:2007vx,castro:2008su,
Castro:2010jq,Fiore:2007vg,Aschieri:2006ye}.
For other ways to conciliate Lorentz Symmetry with noncommutativity
of spacetime see for example \cite{carlson:2002wj,carlson:2002zb,smailagic:2003rp,smailagic:2004yy,ferrari:2006gn}.

Another point of view is to understand Eq.\,(\ref{eq:canonicalNC})
as a first approximation to a more general setting, 
\begin{equation}
\left[\hat{x}_{i}\,,\,\hat{x}_{j}\right]=i\hat{\theta}_{ij}\left(\hat{x}\right)\,,
\end{equation}
where the commutator of coordinates may itself be a non-constant operator,
a function of the coordinates themselves\,\cite{Lukierski:1996br,Gitman:2007ip,Deriglazov:2002gi,Gomes:2009tk,Kupriyanov:2012rf}.
One interesting aspect of this possibility is that one often faces
the appearance of non-Hermitean operators\,\cite{Bagchi:2009wb,Fring:2010pw},
a feature that will somehow appear in our work (for a general discussion
of quantum mechanics with non-Hermitean operators, see the review\,\cite{bender:2007nj}).

Recently, another idea was put forward in\,\cite{Falomir:2009cq},
involving a kind of noncommutativity with mixed spatial and spin degrees
of freedom and a non-relativistic dynamics -- to be hereafter referred
as ``spin noncommutativity''. Such a mixture could be theoretically
understood as a non-relativistic analog of the Snyder's proposal\,(\ref{eq:snydersNC}), where instead of the angular momentum,
the commutator of coordinates is proportional to the spin. Subsequently,
this idea was applied to the study of the Aharonov-Bohm scattering,
which for small distances unveiled a strong anisotropy\,\cite{Das:2011tj}.
In\,\cite{Gomes:2010xk}, the spin noncommutativity was obtained
by means of a consistent deformation of the Berezin-Marinov pseudoclassical
model for the spinning particle\,\cite{Berezin:1976eg}. Besides
that, it was extended to the relativistic situation, and in this context
the spin noncommutativity exhibits at least one advantage over the
canonical one, which is the preservation of the Lorentz symmetry.
Also, a modified Dirac equation for a fermion living in a space with
spin noncommutativity was proposed. 

The aim of the present work is to pursue further the study of the
physical implications of this type of noncommutativity. This work
is organized as follows: our starting point is the noncommutative
Dirac equation defined in\,\cite{Gomes:2010xk}, which is discussed
in Sec.\,\ref{sec:The-noncommutative-Dirac}. The action from which
such equation can be derived is presented and discussed in Sec.\,\ref{sec:The-Action}.
By applying Noether's theorem and also by a direct manipulation of
the equation of motion, we obtain a current which is conserved in
Sec.\,\ref{sec:Conservation-of-the}. Sec.\,\ref{sec:Landau-problem-in}
contains an investigation of the effects of the noncommutativity in
a simple quantum mechanical problem, a particle in the presence of
a constant magnetic field. Here we find that the noncommutative modification
in the theory is embodied in an Hermitean term added to the standard
Dirac Hamiltonian, which is studied perturbatively up to the second
order in the noncommutativity parameter. Finally, Sec.\,\ref{sec:Conclusions-and-Perspectives}
contains our conclusions and perspectives.

\section{\label{sec:The-noncommutative-Dirac}The noncommutative Dirac Equation}

The spin noncommutativity for a relativistic system may be implemented
through the following deformation of the standard position and momentum
operators, 
\begin{equation}
x^{\mu}\rightarrow\hat{x}^{\mu}=x^{\mu}\mathbf{I}+\theta W^{\mu},\,\, p^{\mu}\rightarrow\hat{p}^{\mu}=p^{\mu}\,,\label{eq:BoppShift}
\end{equation}
where $W^{\mu}$ is the Pauli-Lubanski vector
\begin{equation}
W^{\mu}=\frac{1}{2}\varepsilon^{\mu\nu\rho\sigma}p_{\nu}S_{\rho\sigma}=\frac{1}{2}\gamma^{5}\sigma^{\mu\nu}\partial_{\nu}\,,
\end{equation}
and $S_{\rho\sigma}$ is the spin operator. Our conventions are the
following: the flat spacetime metric satisfies $\eta^{00}=-\eta^{ii}=1$,
the Dirac gamma matrices are

\begin{equation}
\gamma^{0}=\left(\begin{array}{cc}
I & 0\\
0 & I
\end{array}\right)\,\,;\,\,\gamma^{i}=\left(\begin{array}{cc}
0 & \sigma^{i}\\
-\sigma^{i} & 0
\end{array}\right)\,,
\end{equation}
in terms of the Pauli matrices $\sigma^{i}$, and also,
\begin{equation}
\sigma^{\mu\nu}=\frac{i}{2}\left(\gamma^{\mu}\gamma^{\nu}-\gamma^{\nu}\gamma^{\mu}\right),\ \ \gamma^{5}=i\gamma^{0}\gamma^{1}\gamma^{2}\gamma^{3}.\,
\end{equation}
A direct consequence of Eq.\,(\ref{eq:BoppShift}) is the noncommutativity
among spacetime coordinates,
\begin{equation}
\left[\hat{x}^{\mu},\hat{x}^{\nu}\right]=-i\theta\varepsilon^{\mu\nu\rho\sigma}S_{\rho\sigma}+i\frac{\theta^{2}}{2}\varepsilon^{\mu\nu\rho\sigma}W_{\rho}p_{\sigma}\,.
\end{equation}
As in Snyder's proposal (\ref{eq:snydersNC}), only covariant objects
appear in this last equation, so Lorentz symmetry is preserved by
this construction.

In standard quantum mechanics, $\hat{x}$ is an observable and therefore
it should necessarily be an Hermitean operator. In our model the position
operator has a non-trivial matrix structure in spinor space, and it
satisfies 
\begin{equation}
\left(\hat{x}^{\mu}\right)^{\dagger}=\gamma^{0}\hat{x}^{\mu}\gamma^{0}.\label{1a}
\end{equation}
The fact that $\hat{x}^{\mu}$ is not Hermitean poses a difficulty
in its interpretation as an observable. One might investigate the
possibility that we are dealing with a $PT$ symmetric system with
a real spectrum for the $\hat{x}^{\mu}$ operator, in which case a
proper redefinition of variables could fix this problem\,\cite{Bagchi:2009wb,Fring:2010pw}.
However, it is far from obvious whether the standard physical interpretation
of the spectra of the coordinate operators applies in a noncommutative
scenario, where exact localization of events in spacetime points is
impossible. In this work, we adopt a more pragmatic point of view,
and we shall consider the commuting coordinate $x^{\mu}$ that will
appear in the noncommutative Dirac equation as a label, in the spirit
of quantum field theory. Besides, we observe that Eq.\,(\ref{1a})
is actually a natural requirement for an operator in spinor space,
which will help to obtain the conjugate Dirac equation and a real
Lagrangian density for our model.

The noncommutative Dirac equation for spin noncommutativity was introduced
in\,\cite{Gomes:2010xk} as 
\begin{equation}
\left\{i\gamma^{\mu}\left[\partial_{\mu}+ieA_{\mu}\left(\hat{x}\right)\right]-m\right\}\psi\left(x\right)=0,
\end{equation}
where the operator $A_{\mu}\left(\hat{x}\right)$ is constructed from
$\hat{x}$ via the Weyl (symmetric) ordering, 
\begin{equation}
f\left(\hat{x}\right)=\int\frac{d^{4}k}{\left(2\pi\right)^{4}}\,\tilde{f}\left(k\right)\, e^{-ik_{\mu}\hat{x}^{\mu}}\,.
\end{equation}
It should be noted that the operator $A_{\mu}\left(x^{\mu}\mathbf{I}+\theta\gamma^{5}\sigma^{\mu\nu}\partial_{\nu}\right)$
has a nontrivial matrix structure which does not commute with $\gamma^{\mu}$,
so we face an ordering ambiguity in the noncommutative generalization
for the matrix product $\gamma^{\mu}A_{\mu}$. By introducing both
left and right orderings with arbitrary coefficients $a_{1}$ and
$a_{2}$ such that $a_{1}+a_{2}=1$, we may define a deformed Dirac
equation 
\begin{equation}
\hat{O}\psi=\left[i\gamma^{\mu}\partial_{\mu}-m-e\left(a_{1}\gamma^{\mu}A_{\mu}\left(\hat{x}\right)+a_{2}A_{\mu}\left(\hat{x}\right)\gamma^{\mu}\right)\right]\psi\left(x\right)=0\,.\label{eq:NCDirac1}
\end{equation}
For the ordinary commutative spacetime, the property 
\begin{equation}
\left[i\gamma^{\mu}\partial_{\mu}-m-e\gamma^{\mu}A_{\mu}\right]^{\dagger}=\gamma^{0}\left[i\gamma^{\mu}\partial_{\mu}-m-e\gamma^{\mu}A_{\mu}\right]\gamma^{0}\,,
\end{equation}
may be used to prove that the Hamiltonian operator one finds when writing
the ordinary Dirac equation in the form $i\partial_{t}\psi=H\psi$
is Hermitean. It is therefore natural to require that the operator
$\hat{O}$ appearing in Eq. (\ref{eq:NCDirac1}) also satisfies
\begin{equation}
\hat{O}^{+}=\gamma^{0}\hat{O}\gamma^{0}\,.\label{eq:DiracConj}
\end{equation}
In the noncommutative case, one has to demand that $a_{1}=a_{2}$,
i.e., \emph{symmetric ordering}, to have this property. The symmetric
ordering will also ensure the reality of the Lagrangean density corresponding
to Eq.\,(\ref{eq:NCDirac1}); finally it simplifies considerably
the derivation of the noncommutative Hamiltonian we will discuss in
Sec.\,\ref{sec:Landau-problem-in}. These facts are enough
for us to choose hereafter the ordering defined by $a_{1}=a_{2}=1/2$,
which fixes the noncommutative form of the Dirac equation as
\begin{equation}
\left[i\gamma^{\mu}\partial_{\mu}-m-\frac{e}{2}\left(\gamma^{\mu}A_{\mu}\left(\hat{x}\right)+A_{\mu}\left(\hat{x}\right)\gamma^{\mu}\right)\right]\psi\left(x\right)=0\,.\label{eq:NCDirac2}
\end{equation}
An interesting feature of this model is that, in spite of the presence
of noncommutativity and nonlocality, it is Lorentz invariant, in the
sense that the deformed Dirac equation in Eq.\,(\ref{eq:NCDirac2})
is Lorentz covariant, and the noncommutative parameter $\theta$ is
a Lorentz scalar. Of course, the defining map in Eq.\,(\ref{eq:BoppShift})
was devised for this to happen, since it only contains covariant objects.

The action of Weyl ordered operator $f\left(x^{\mu}\mathbf{I}+\theta\gamma^{5}\sigma^{\mu\nu}\partial_{\nu}\right)$
on a spinor $\psi\left(x\right)$ can be represented by means of a
``star operation'' $\star$ as follows,
\begin{align}
f\left(x^{\mu}\mathbf{I}+\theta\gamma^{5}\sigma^{\mu\nu}\partial_{\nu}\right)\psi & =f\star\psi=f\exp\left(\theta\overleftarrow{\partial_{\mu}}\gamma^{5}\sigma^{\mu\nu}\overrightarrow{\partial_{\nu}}\right)\psi\,,\label{eq:starop1}
\end{align}
or, more explictly,
\begin{equation}
f\star\psi=f\psi+\theta\partial_{\mu}f\gamma^{5}\sigma^{\mu\nu}\partial_{\nu}\psi+\frac{\theta^{2}}{2}\partial_{\mu_{1}}\partial_{\mu_{2}}f\gamma^{5}\sigma^{\mu_{1}\nu_{1}}\gamma^{5}\sigma^{\mu_{2}\nu_{2}}\partial_{\nu_{1}}\partial_{\nu_{2}}\psi+\ldots\,.\label{eq:staropexpl}
\end{equation}
We note that the star operation defined above involves a regular (scalar)
function $f$ and a Dirac spinor (column vector): it is not a ``star
product'' in the usual sense since it does not map two similar objects
in the same class of objects, therefore we cannot even discuss associativity
of this operation. We will shortly define what we mean by a ``star
operation'' involving other objects such as conjugate spinors, and
then discuss some of its properties. The relevant fact at this point
is that the noncommutative Dirac equation\,(\ref{eq:NCDirac2}) can
be cast in terms of the star operation as
\begin{equation}
\left[-i\gamma^{\mu}\partial_{\mu}+m+\frac{e}{2}\left(A_{\mu}\left(x\right)\gamma^{\mu}\star+A_{\mu}\left(x\right)\star\gamma^{\mu}\right)\right]\psi\left(x\right)=0\,,\label{eq:EqDiracStar}
\end{equation}
which, taking into account Eq.\,(\ref{eq:staropexpl}), turns out
to be
\begin{multline}
\left[-i\gamma^{\mu}\partial_{\mu}+m+e\gamma^{\mu}A_{\mu}\left(x\right)+\frac{e\theta}{2}\partial_{\alpha_{1}}A_{\mu}\left(\gamma^{\mu}\gamma^{5}\sigma^{\alpha_{1}\beta_{1}}+\gamma^{5}\sigma^{\alpha_{1}\beta_{1}}\gamma^{\mu}\right)\partial_{\beta_{1}}\right.\\
\left.+\frac{e\theta^{2}}{4}\partial_{\alpha_{1}}\partial_{\alpha_{2}}A_{\mu}\left(\gamma^{\mu}\sigma^{\alpha_{1}\beta_{1}}\sigma^{\alpha_{2}\beta_{2}}+\sigma^{\alpha_{1}\beta_{1}}\sigma^{\alpha_{2}\beta_{2}}\gamma^{\mu}\right)\partial_{\beta_{1}}\partial_{\beta_{2}}+...\right]\psi\left(x\right)\,=\,0\,.\label{eq:EqDiracExpanded}
\end{multline}
The noncommutative Dirac equation has in general
an infinite tower of time derivatives, so the usual Hamiltonian interpretation
of quantum mechanics -- based on an Hermitean Hamiltonian, which ensures
unitarity of time evolution and conservation of probability -- is
not possible. Inspite of that, we shall demonstrate in Sec.\,\ref{sec:Conservation-of-the}
that a conserved charge current can be defined in general, and for
a particular choice of $A_{\mu}$, we shall be able to derive a consistent
Hamiltonian formulation in Sec.\,\ref{sec:Landau-problem-in}.

\section{\label{sec:The-Action}The Action}

The star operation and its properties are useful in deriving the noncommutative
Dirac equation\,(\ref{eq:NCDirac2}) from an action principle. We
start by obtaining the conjugate Dirac equation, and for that end
we define a star operation between the usual Dirac conjugate spinor
$\bar{\psi}=\psi^{\dagger}\gamma^{0}$ and a function$f$ by the following
rule,
\begin{equation}
\bar{\psi}\star f\,\equiv\,\left(f\star\psi\right)^{\dagger}\gamma^{0}\,,
\end{equation}
or, more explictly,
\begin{align}
\bar{\psi}\star f= & \bar{\psi}\exp\left(\theta\overleftarrow{\partial_{\mu}}\gamma^{5}\sigma^{\mu\nu}\overrightarrow{\partial_{\nu}}\right)f\label{eq:starop2}\\
= & \bar{\psi}f+\theta\partial_{\mu}\bar{\psi}\gamma^{5}\sigma^{\mu\nu}\partial_{\nu}f+\mathcal{O}\left(\theta^{2}\right)\,.\nonumber 
\end{align}
Finally, we introduce a star operation between two spinors by the
formula
\begin{align}
\bar{\varphi}\star\psi & =\bar{\varphi}\exp\left(\theta\overleftarrow{\partial_{\mu}}\gamma^{5}\sigma^{\mu\nu}\overrightarrow{\partial_{\nu}}\right)\psi
\label{eq:starop3}\\
 & =\bar{\varphi}\psi+\theta\partial_{\mu}\bar{\varphi}\gamma^{5}\sigma^{\mu\nu}\partial_{\nu}\psi+\mathcal{O}\left(\theta^{2}\right)
 \nonumber \,.
\end{align}

We may now quote some useful properties that can be proved regarding
the star operation defined in Eqs.\,(\ref{eq:starop1}),\,(\ref{eq:starop2})
and\,(\ref{eq:starop3}). First of all, we find the exact equality
\begin{equation}
\left(\bar{\varphi}\star\psi\right)^{*}=\bar{\psi}\star\varphi\,.
\end{equation}
Next, integration by parts and the antisymmetry of $\sigma^{\mu\nu}$
leads to 
\begin{equation}
\int d^{4}x\,\bar{\varphi}\star\psi=\int d^{4}x\,\bar{\varphi}\psi+\mbox{surface terms}\,,\label{eq:prop1}
\end{equation}
a property that is well known from the studies involving canonical
noncommutativity and its associated star (Moyal) product.

We will also need to manipulate expressions of the general form $\int d^{4}x\,\bar{\varphi}\left(f\star\psi\right)$,
involving two arbitrary spinors $\varphi$ and $\psi$ and a function
$f$. Starting with
\begin{align}
\int d^{4}x\bar{\varphi}\left(f\star\psi\right) & =\int d^{4}x\Big(\bar{\varphi}f\psi+\theta\bar{\varphi}\partial_{\mu}f\gamma^{5}\sigma^{\mu\nu}\partial_{\nu}\psi\nonumber \\
 & +\frac{\theta^{2}}{2}\bar{\varphi}\partial_{\mu_{1}}\partial_{\mu_{2}}f\gamma^{5}\sigma^{\mu_{1}\nu_{1}}\gamma^{5}\sigma^{\mu_{2}\nu_{2}}\partial_{\nu_{1}}\partial_{\nu_{2}}\psi+...\Big)\,,
\end{align}
one integrates by parts all derivatives acting on $\psi$, taking
care of the antisymmetry of $\sigma^{\alpha\beta}$, obtaining 
\begin{align}
\int d^{4}x\bar{\varphi}\left(f\star\psi\right)= & \int d^{4}x \Big[\bar{\varphi}f\psi+\theta\partial_{\mu}\bar{\varphi}\partial_{\nu}f\gamma^{5}\sigma^{\mu\nu}\psi \nonumber \\
 & +\frac{\theta^{2}}{2}\partial_{\nu_{1}}\partial_{\nu_{2}}\bar{\varphi}\partial_{\mu_{1}}\partial_{\mu_{2}}f\gamma^{5}\sigma^{\mu_{1}\nu_{1}}\gamma^{5}\sigma^{\mu_{2}\nu_{2}}\psi+\cdots\nonumber \\
 & +\partial_{\mu}E^{\mu}\Big]\,,\label{eq:prop2a}
\end{align}
where
\begin{align}
E^{\mu} \,=\, & \theta\bar{\varphi}\partial_{\nu}f\gamma^{5}\sigma^{\nu\mu}\psi+\frac{\theta^{2}}{2}\bar{\varphi}\partial_{\mu_{1}}\partial_{\mu_{2}}f\gamma^{5}\sigma^{\mu_{1}\nu_{1}}\gamma^{5}\sigma^{\mu_{2}\mu}\partial_{\nu_{1}}\psi\nonumber \\
 & -\frac{\theta^{2}}{2}\partial_{\nu_{2}}\bar{\varphi}\partial_{\mu_{1}}\partial_{\mu_{2}}f\gamma^{5}\sigma^{\mu_{1}\mu}\gamma^{5}\sigma^{\mu_{2}\nu_{2}}\psi+\mathcal{O}\left(\theta^{3}\right)\,.
\end{align}
Then one recognizes in the right hand side of (\ref{eq:prop2a}) the
expansion of $\left(\bar{\varphi}\star f\right)\psi$, i.e.,
\begin{equation}
\int d^{4}x\bar{\varphi}\left(f\star\psi\right)=\int d^{4}x\left[\left(\bar{\varphi}\star f\right)\psi+\partial_{\mu}E^{\mu}\right]\,.\label{eq:prop2}
\end{equation}
It should be stressed that, while we have only explicitly written
$E^{\mu}$ up to the second order in $\theta$, the fact that Eq.
(\ref{eq:prop2}) holds (i.e., the difference between the two integrals
is a surface term) actually is true for any order of $\theta$, as
it is clear from this derivation.

In particular, expressions like the one in Eq.\,(\ref{eq:prop2})
will appear in which $f$ is the electromagnetic potential $A_{\mu}$,
which always appears contracted with a $\gamma^{\mu}$. In this case,
one should be careful with the order of the star operation and the
$\gamma^{\mu}$ since they do not commute. In any case, it can be
shown that,
\begin{subequations}\label{eq:prop3a}
\begin{eqnarray}
\int d^{4}x\,\bar{\varphi}\left(A_{\mu}\star\gamma^{\mu}\psi\right) & = & \int d^{4}x\left[\left(\bar{\varphi}\star A_{\mu}\gamma^{\mu}\right)\psi+\partial_{\mu}F^{\mu}\right]\,,\\
\int d^{4}x\,\bar{\varphi}\left(A_{\mu}\gamma^{\mu}\star\psi\right) & = & \int d^{4}x\left[\left(\bar{\varphi}\gamma^{\mu}\star A_{\mu}\right)\psi+\partial_{\mu}G^{\mu}\right]\,,
\end{eqnarray}
\end{subequations}
where\begin{subequations}\label{eq:prop3b}
\begin{align}
G^{\mu} & =\theta\bar{\varphi}\partial_{\alpha}A_{\nu}\gamma^{5}\sigma^{\alpha\mu}\gamma^{\nu}\psi+\mathcal{O}\left(\theta^{2}\right)\,,\\
H^{\mu} & =\theta\bar{\varphi}\partial_{\alpha}A_{\nu}\gamma^{\nu}\gamma^{5}\sigma^{\alpha\mu}\psi+\mathcal{O}\left(\theta^{2}\right)\,.
\end{align}
\end{subequations}

Finally, we can write an action describing the interaction of a Dirac
fermion with an electromagnetic potential $A_{\mu}$ in a spacetime
with spin noncommutativity,
\begin{equation}
S\left[\psi,A\right]=\int d^{4}x\,\bar{\psi}\left(x\right)\left[-i\gamma^{\mu}\partial_{\mu}\psi\left(x\right)+m\psi\left(x\right)
+\frac{e}{2}\left(
A_{\mu}\left(x\right)\gamma^{\mu}\star+A_{\mu}\left(x\right)\star\gamma^{\mu}\right)\psi\left(x\right) \right]
\,.\label{eq:NCDiracAction}
\end{equation}
Clearly, Eq.\,(\ref{eq:EqDiracStar}) is obtained by variation of
Eq.\,(\ref{eq:NCDiracAction}), $\delta S/\delta\bar{\psi}\left(x\right)=0$.
We split, as usual, this action in free and interaction part,
\begin{equation}
S=S_{0}+S_{I}\,.\label{13}
\end{equation}
The usual free Dirac action
\begin{align}
S_{0}= & \int d^{4}x\,\bar{\psi}\left(-i\gamma^{\mu}\partial_{\mu}\psi+m\psi\right)\,,\label{eq:S0}
\end{align}
could also be written in a more symmetrical form involving the star
operation due to Eq.\,(\ref{eq:prop1}),
\[
S_{0}=\int d^{4}x\left(-\frac{i}{2}\bar{\psi}\star\gamma^{\mu}\partial_{\mu}\psi+\frac{i}{2}\partial_{\mu}\bar{\psi}\gamma^{\mu}\star\psi+m\bar{\psi}\star\psi\right)\,.
\]
On the other hand, for the interaction part we write
\begin{align}
S_{I}= & \frac{e}{2}\int d^{4}x\,\bar{\psi}\left(\gamma^{\mu}A_{\mu}\star\psi+A_{\mu}\star\gamma^{\mu}\psi\right)\label{eq:Sint}\\
= & \frac{e}{4}\int d^{4}x\left[\bar{\psi}\left(\gamma^{\mu}A_{\mu}\star\psi\right)+\left(\bar{\psi}\gamma^{\mu}\star A_{\mu}\right)\psi+\bar{\psi}\left(A_{\mu}\star\gamma^{\mu}\psi\right)+\left(\bar{\psi}\star\gamma^{\mu}A_{\mu}\right)\psi\right]\,,
\end{align}
where we have used Eq.\,(\ref{eq:prop3a}), and finally
\begin{align}
S_{I}= & \frac{e}{4}\int d^{4}x\left[\bar{\psi}\star\left(\gamma^{\mu}A_{\mu}\star\psi\right)+\left(\bar{\psi}\gamma^{\mu}\star A_{\mu}\right)\star\psi\right.\nonumber \\
 & \left.+\bar{\psi}\star\left(A_{\mu}\star\gamma^{\mu}\psi\right)+\left(\bar{\psi}\star\gamma^{\mu}A_{\mu}\right)\star\psi\right]\,,
\end{align}
after using Eq.\,(\ref{eq:prop1}).

One can verify that the action in Eq.\,(\ref{eq:NCDiracAction})
is real. This property is a consequence of Eq.\,(\ref{eq:DiracConj}),
which is only valid if we adopt the symmetric ordering as in Eq.\,(\ref{eq:NCDirac2}).

Finally, one might use the properties of the star operation quoted in this Section to show that
the equation satisfied by the Dirac conjugate $\bar{\psi}$ reads
\begin{equation}
\bar{\psi}\left(x\right)\left[i\gamma^{\mu}\overleftarrow{\partial}_{\mu}+m+\frac{e}{2}\star\gamma^{\mu}A_{\mu}\left(x\right)+\frac{e}{2}\gamma^{\mu}\star A_{\mu}\left(x\right)\right]=0\,.\label{eq:DiracEqConj}
\end{equation}
Then, Eqs.\,(\ref{eq:prop1}) and\,(\ref{eq:prop3a})
are used to rewrite Eq.\,(\ref{eq:NCDiracAction}) in the form
\begin{equation}
S=\int d^{4}x\left[i\partial_{\mu}\bar{\psi}\gamma^{\mu}+m\bar{\psi}+\frac{e}{2}\left(\bar{\psi}\star A_{\mu}\gamma^{\mu}+\bar{\psi}\gamma^{\mu}\star A_{\mu}\right)\right]\psi\,,
\end{equation}
and now the variation over $\psi\left(x\right)$ gives the conjugate
Dirac equation in Eq.\,(\ref{eq:DiracEqConj}), as it should be.

\section{\label{sec:Conservation-of-the}Conservation of the Electrical Current}

In this section, we want to find an expression for the conserved electric
current $j^{\mu}$ in our theory, since the existence of such a current
is crucial for the physical meaning of the model. The action (\ref{eq:NCDiracAction})
has global phase invariance, so Noether's theorem provides a general
formula for the associated conserved current. Due to the appearance
of arbitrary high-order derivatives in $\psi$, one would need to
generalize the well-known formula for the Noether current (see for
example~\cite{Gomes:2011di}). Expanding Eq.\,(\ref{eq:Sint}) in
the first order of $\theta$, however, one finds
\begin{equation}
S_{I}=e\int d^{4}x\, A_{\mu}\bar{\psi}\gamma^{\mu}\psi-\frac{e\theta}{2}\int d^{4}x\,\bar{\psi}\gamma^{5}\left[\sigma^{\mu\nu},\gamma^{\alpha}\right]\partial_{\mu}A_{\alpha}\partial_{\nu}\psi+\mathcal{O}\left(\theta^{2}\right)\,,
\end{equation}
which, with the help of the identities
\begin{subequations}\label{eq:sigma_gamma}
\begin{align}
\gamma^{\mu}\sigma^{\rho\sigma} & =+\left(\eta^{\rho\mu}\gamma^{\sigma}-\eta^{\sigma\mu}\gamma^{\rho}\right)-i\varepsilon^{\mu\nu\rho\sigma}\gamma^{5}\gamma_{\nu}\,,\\
\sigma^{\rho\sigma}\gamma^{\mu} & =-\left(\eta^{\rho\mu}\gamma^{\sigma}-\eta^{\sigma\mu}\gamma^{\rho}\right)-i\varepsilon^{\mu\nu\rho\sigma}\gamma^{5}\gamma_{\nu}\,,
\end{align}
\end{subequations}
can be cast as
\begin{equation}
S_{I}=\int d^{4}x\, eA_{\mu}\bar{\psi}\gamma^{\mu}\psi-ie\theta\int d^{4}x\,\bar{\psi}\gamma^{5}\left(\partial_{\mu}A^{\nu}\gamma^{\mu}\partial_{\nu}\psi-\partial_{\mu}A^{\mu}\gamma^{\nu}\partial_{\nu}\psi\right)+\mathcal{O}\left(\theta^{2}\right)\,.
\end{equation}
Since, in this approximation, there are no higher-order derivatives
acting on $\psi$, on may use the standard formula for the Noether
current associated to the phase symmetry $\delta\psi=-i\alpha\psi$,
\begin{align}
j^{\mu} & =-i\frac{\partial\mathcal{L}}{\partial\left(\partial_{\mu}\psi\right)}\psi\\
 & =\bar{\psi}\gamma^{\mu}\psi+e\theta\bar{\psi}\gamma^{5}\left(\partial_{\nu}A^{\nu}\gamma^{\mu}-\partial_{\nu}A^{\mu}\gamma^{\nu}\right)\psi+\mathcal{O}\left(\theta^{2}\right)\,.\label{eq:current1}
\end{align}

To see the existence of a conserved current $j^{\mu}$ at arbitrary
order in $\theta$, we shall employ the following trick: using Eqs.
(\ref{eq:NCDirac2},\ref{eq:DiracEqConj}) one may write the identity,
\begin{eqnarray}
\left(i\partial_{\mu}\bar{\psi}\gamma^{\mu}+m\bar{\psi}+\frac{e}{2}\bar{\psi}\star\gamma^{\mu}A_{\mu}+\frac{e}{2}\bar{\psi}\gamma^{\mu}\star A_{\mu}\right)\psi\nonumber \\
-\bar{\psi}\left(-i\gamma^{\mu}\partial_{\mu}\psi+m\psi+\frac{e}{2}\gamma^{\mu}A_{\mu}\star\psi+\frac{e}{2}A_{\mu}\star\gamma^{\mu}\psi\right) & = & 0\,.\label{eq:trick1}
\end{eqnarray}
In the usual case (without the star operation), all that would remain
would be $i\partial_{\mu}\bar{\psi}\gamma^{\mu}\psi+i\bar{\psi}\gamma^{\mu}\partial_{\mu}\psi=\partial_{\mu}\left(i\bar{\psi}\gamma^{\mu}\psi\right)=0$,
 giving the conservation of the usual electric current. In our
case, Eq.\,(\ref{eq:trick1}) can be written as
\begin{align}
\partial_{\mu}\left(\bar{\psi}\gamma^{\mu}\psi\right) & +i\frac{e}{2}\left[\left(\bar{\psi}\star\gamma^{\mu}A_{\mu}\right)\psi-
\bar{\psi}\left(A_{\mu}\star\gamma^{\mu}\psi\right)\right]\nonumber \\
 & +i\frac{e}{2}\left[\left(\bar{\psi}\gamma^{\mu}\star A_{\mu}\right)\psi
-\bar{\psi}\left(\gamma^{\mu}A_{\mu}\star\psi\right)\right]=0\,,\label{eq:trick2}
\end{align}
which, by virtue of Eqs.\,(\ref{eq:prop3a},\ref{eq:prop3b}), leads
to
\begin{equation}
\partial_{\mu}\left[\bar{\psi}\gamma^{\mu}\psi+
i\frac{e\theta}{2}\bar{\varphi}\partial_{\alpha}A_{\nu}\left(\gamma^{5}\sigma^{\alpha\mu}\gamma^{\nu}
+\gamma^{\nu}\gamma^{5}\sigma^{\alpha\mu}\right)\psi +\mathcal{O}\left(\theta^{2}\right) \right]=0\,.
\end{equation}
Finally, from Eqs.\,(\ref{eq:sigma_gamma}),
\begin{equation}
\partial_{\mu}\left[\bar{\psi}\gamma^{\mu}\psi+e\theta\bar{\psi}\gamma^{5}\left(\partial_{\nu}A^{\nu}\gamma^{\mu}
-\partial_{\nu}A^{\mu}\gamma^{\nu}\right)\psi+\mathcal{O}\left(\theta^{2}\right)\right]=0\,\label{eq:trick3}
\end{equation}
For the reasons commented in the paragraph containing Eq.\,(\ref{eq:prop2}),
this last equation holds for any order of $\theta$, which ensures
the existence of the conserved current $j^{\mu}$.

It is noteworthy that the conserved current depends, already at first
order in $\theta$, on the electromagnetic potential, which we have
treated as a fixed background field. The same feature appears in canonical
noncommutativity\,\cite{Adorno:2011wj}, and it makes
interesting the problem of incorporating a dynamical
potential $A_{\mu}$ in a consistent way.

\section{\label{sec:Landau-problem-in}Landau problem in the presence of spin
noncommutativity}

Having further explored the formal aspects of the spin noncommutativity,
in this section we want to gain some insight into its possible observable
consequences in a particular physical problem. We consider the bound
state problem for a charged particle subject to a constant magnetic
field, known as the Landau problem. 

When the noncommutativity is not present, the Landau problem is described
in many textbooks such as\,\cite{itzykson1980quantum}. The gauge
potential corresponding to a constant magnetic field perpendicular
to the $xy$ plane can be chosen as 
\begin{equation}
A^{2}=Bx^{1}\qquad\mbox{and}\qquad A^{0}=A^{1}=A^{3}=0\,,
\end{equation}
and the Dirac's Hamiltonian 
\begin{equation}
H_{0}=-i\gamma^{0}\gamma^{i}\partial_{i}+m\gamma^{0}+eBx_{1}\gamma^{0}\gamma^{2}\,,\label{eq:H0}
\end{equation}
has energy levels 
\[
E_{n,\alpha}=\sqrt{p_{3}^{2}+m^{2}+eB\left(2n+1-\alpha\right)}\,,
\]
with $\alpha=\pm1$ for spin up and down, respectively. All energy
levels exhibit an infinite-degeneracy relative to $p_{1}$ and $p_{2}$.
Besides that, except for the (unique) ground-state $\left|0\right\rangle =\left|0,+1\right\rangle $,
the excited energy levels are two-fold degenerate, since $\left|n,+1\right\rangle $
and $\left|n-1,-1\right\rangle $ have the same energy. The energy
eigenfunctions of $H_{0}$ can be cast in terms of the two-components
eigenvectors $\chi_{\alpha}$ of the Pauli matrix $\sigma^{3}$,
\begin{equation}
\sigma^{3}\chi_{\alpha}=\alpha\chi_{\alpha}\,,
\end{equation}
as follows,
\begin{equation}
\left|n,\alpha\right\rangle =c_{n,\alpha}\left(\begin{array}{c}
\left|n\right\rangle \chi_{\alpha}\\
\frac{\vec{\sigma}\cdot\vec{\pi}}{E_{n,\alpha}+m}\left|n\right\rangle \chi_{\alpha}
\end{array}\right)\,.\label{eq:eigenfunctions}
\end{equation}
Here, $\left|n\right\rangle $ are essentially eigenstates of the
harmonic oscillator, 
\begin{equation}
\varphi_{n}\left(x\right)=\left\langle x|n\right\rangle ={\rm e}^{i\left(p_{2}x_{2}+ip_{3}x_{3}\right)}e^{-\xi^{2}/2}H_{n}\left(\xi\right)\,,
\end{equation}
where
\begin{equation}
\xi=\sqrt{eB}\left(x_{2}-\frac{p_{2}}{eB}\right)\,,
\end{equation}
and $c_{n,\alpha}$ is a normalization factor,
\begin{equation}
c_{n,\alpha}=\frac{\left(eB\right)^{1/4}}{2\pi}\sqrt{\frac{E_{n,\alpha}+m}{E_{n,\alpha}}}\frac{1}{\sqrt{\sqrt{\pi}2^{n+1}n!}}\,.
\end{equation}
Our choice of $c_{n,\alpha}$ is slightly different from the usual
one, but it has the advantage that the eigenfunctions in Eq.\,(\ref{eq:eigenfunctions})
are orthonormal in the simplest sense, i.e.,
\begin{equation}
\left\langle n^{\prime},\alpha^{\prime}|n,\alpha\right\rangle =\delta_{n^{\prime},n}\delta_{\alpha^{\prime},\alpha}\,.
\end{equation}
The canonical momentum $\vec{\pi}$ is
\begin{equation}
\vec{\pi}\equiv\left(-i\partial_{1},p_{2}-eBx_{1},p_{3}\right)=\sqrt{eB}\left(-i\partial_{\xi},-\xi,\frac{p_{3}}{\sqrt{eB}}\right)\,,
\end{equation}
and we quote some formulae that are useful in calculating matrix elements
of the deformed Hamiltonian,\begin{subequations}\label{eq:ortho}
\begin{align}
\pi^{1}\left(c_{n,\alpha}\left|n\right\rangle \right) & =i\sqrt{eB}\left(\frac{1}{2}\frac{c_{n,\alpha}}{c_{n+1,\alpha}}\left|n+1\right\rangle -\frac{nc_{n,\alpha}}{c_{n-1,\alpha}}\left|n-1\right\rangle \right)\,,\\
\pi^{2}\left(c_{n,\alpha}\left|n\right\rangle \right) & =\hphantom{i}\sqrt{eB}\left(\frac{1}{2}\frac{c_{n,\alpha}}{c_{n+1,\alpha}}\left|n+1\right\rangle +\frac{nc_{n,\alpha}}{c_{n-1,\alpha}}\left|n-1\right\rangle \right)\,,\\
\pi^{3}\left|n\right\rangle  & =p_{3}\left|n\right\rangle \,.
\end{align}
\end{subequations}

The fact that the noncommutative Dirac equation (\ref{eq:EqDiracStar})
has in general higher orders in time derivatives precludes the definition
of a Dirac Hamiltonian in the standard way. It is actually a consequence
of Lorentz invariance that the non-locality in space introduced by
noncommutativity should also extend to the time variable. This difficulty
is circumvented in the particular problem studied in this section
because the linearity of the electromagnetic potential makes the noncommutativity
modification of the Dirac equation local both in time and space, and
all higher orders corrections in Eq.\,(\ref{eq:EqDiracExpanded})
vanish. We end up with with the simple Hamiltonian 
\begin{equation}
H=H_{0}+H_{I}\,,\label{eq:HamiltonianMod}
\end{equation}
where
\begin{align}
H_{I} & =\frac{i}{2}\theta\, eB\, p_{2}\gamma^{2}\gamma^{3}\nonumber \\
 & =-\theta\,\frac{eB}{2}\, p_{2}\left(\begin{array}{cc}
\sigma^{1} & 0\\
0 & \sigma^{1}
\end{array}\right)\,.\label{eq:Hnc}
\end{align}
It should be stressed that Eq.\,(\ref{eq:HamiltonianMod}) contains
the exact modification of the Hamiltonian for the present problem.
This observation is necessary since we will use the $\mathcal{O}\left(\theta\right)$
correction in Eq.\,(\ref{eq:HamiltonianMod}) to calculate the corrections
to the energy levels up to $\mathcal{O}\left(\theta^{2}\right)$ in
the sequel. Another remark is that the symmetric ordering adopted
in Eq.\,(\ref{eq:NCDirac2}) is also essential in keeping the noncommutative
modification to the Hamiltonian exactly of first order in $\theta$:
if we had $a_{1}-a_{2}\neq0$, the calculation of the noncommutative
Hamiltonian would involve a multiplicative factor $\left[\gamma^{0}-\frac{i}{2}\left(a_{1}-a_{2}\right)\theta eB\gamma^{3}\right]^{-1}$,
which would introduce higher orders corrections. One may also quickly
verify that Eq.\,(\ref{eq:Hnc}) is Hermitean, so it maintains the
reality of the energy spectrum.

First order corrections to the energy of the ground-state $\left|0\right\rangle =\left|0,+1\right\rangle $
are given by
\begin{equation}
\delta E_{0}^{\left(1\right)}=\left\langle 0\left|H_{I}\right|0\right\rangle =-\frac{eB\theta}{2}p_{2}\left\langle 0\left|\left(\begin{array}{cc}
\sigma^{1} & 0\\
0 & \sigma^{1}
\end{array}\right)\right|0\right\rangle \,.
\end{equation}
Using the orthogonality relations\,(\ref{eq:ortho}), as well standard
Dirac matrices manipulations, one can show that
\begin{equation}
\delta E_{0}^{\left(1\right)}=0\,.
\end{equation}
For the correction to the degenerate energy levels, 
\begin{equation}
E_{n}=\sqrt{p_{3}^{2}+m^{2}+2neB}\,,\label{eq:degenerateenergies}
\end{equation}
with $n\geq1$, one has to solve the secular equation
\begin{equation}
\mbox{det}\left(W_{ij}-\delta E_{n}^{\left(1\right)}\delta_{ij}\right)=0\,,\label{eq:sec1}
\end{equation}
where
\begin{equation}
W_{ij}=\left\langle n,i\left|H_{I}\right|n,j\right\rangle \,.\label{eq:sec1a}
\end{equation}
Here, $\left|n,1\right\rangle =\left|n,+1\right\rangle $ and $\left|n,2\right\rangle =\left|n-1,-1\right\rangle $
is a basis for each degenerate level. Calculation of the matrix elements
$W_{ij}$ is also straightforward, and one can show that $W_{ij}=0$,
so that
\begin{equation}
\delta E_{n}^{\left(1\right)}=0\,.
\end{equation}
In summary, the spin noncommutativity does not change the spectrum
of the Landau problem in the first order in $\theta$.

We found non-trivial corrections to the energy levels in the second
order of perturbation theory. For the ground-state energy, one has
to calculate
\begin{equation}
\delta E_{0}^{\left(2\right)}=\sum_{n,i}\frac{\left|W_{n,i}\right|^{2}}{E_{0}-E_{n}}\quad n\geq1\mbox{ and }i=1,2\,,
\end{equation}
where $W_{n,i}$ are matrix elements of $H_{I}$ between the ground-state
and the excited state $\left|n,i\right\rangle $. The only nonvanishing
of these matrix elements are\begin{subequations}
\begin{align}
W_{1,1} & =\frac{i\theta\left(Be\right)^{3/2}p_{2}p_{3}}{2\left(E_{0}+m\right)\left(E_{1}+m\right)}\,\frac{c_{0,+1}}{c_{1,+1}}\,,\\
W_{1,2} & =\frac{\theta\, eB\, p_{2}}{2}\frac{\left[c_{1,-1}c_{1,+1}\left(2eB+p_{3}^{2}\right)+c_{0,-1}c_{0,+1}eB\right]}{c_{1,-1}c_{1,+1}\left(E_{0}+m\right)\left(E_{1}+m\right)}\,,\\
W_{3,2} & =\frac{3\theta\left(eB\right)^{2}p_{2}}{\left(E_{0}+m\right)\left(E_{3}+m\right)}\,\frac{c_{0,+1}c_{2,-1}}{c_{1,-1}c_{1,+1}}\,,
\end{align}
\end{subequations}from which the final (nonvanishing) expression
for $\delta E_{0}^{\left(2\right)}$ can be calculated,
\begin{align}
\delta  E_{0}^{\left(2\right)}= & -\frac{\theta^{2}\left(eB\right)^{2}p_{2}^{2}}{4\left(E_{0}+m\right)^{2}}\left[\frac{eB\, p_{3}^{2}}{\left(E_{1}-E_{0}\right)\left(E_{1}+m\right)^{2}}\left(\frac{c_{0,+1}}{c_{1,+1}}\right)^{2}+\right.\nonumber \\
 & +\frac{1}{\left(E_{1}-E_{0}\right)\left(E_{1}+m\right)^{2}}\left(eB\left(\frac{c_{0,+1}c_{0,-1}}{c_{1,-1}c_{1,+1}}-2\right)+p_{3}\right)^{2}\nonumber \\
 & \left.+\frac{\left(eB\right)^{2}p_{3}^{2}}{\left(E_{3}-E_{0}\right)\left(E_{3}+m\right)^{2}}\left(\frac{c_{0,+1}c_{2,-1}}{c_{1,+1}c_{1,-1}}\right)^{2}\right]\,.
\end{align}

More interesting is the calculation of the second-order energy corrections
to the degenerate levels, since there we can investigate whether the
degeneracy is broken by the noncommutativity. Physically, when the
perturbation breaks the degeneracy, that means some symmetry is broken;
in our problem, it is the constant magnetic field which breaks part
of the rotational symmetry. In the commutative case,
one still has the two-fold degeneracy of the excited levels $\left|n,i\right\rangle $.
Since the noncommutative correction to the Hamiltonian $H_{I}$ depends
on the magnetic field, it might be that this degeneracy is broken,
even if the noncommutativity itself does not break further symmetries. 

Second order corrections to the energy of degenerate levels are found
by solving the secular equation\,\cite{LandauQMNR}
\begin{equation}
\mbox{det}\left(W_{ij}+\sum\frac{W_{n,i;m,\ell}W_{m,\ell;n,j}}{E_{n}-E_{m}}-\delta E_{n}^{\left(2\right)}\delta_{ij}\right)=0\,,\label{eq:sec2}
\end{equation}
where the sum is for $m\geq1$ and $m\neq n$, and $\ell=1,2$, and
$W_{n,i;m,\ell}$ is the matrix element of $H_{I}$ between two degenerate
states. This calculation is straightforward but quite involved, so
it was done using a Computer Algebra System (CAS)\,\cite{wolfram:2003}.
The resulting expressions are too long and not particularly informative to be
quoted here, but the
relevant fact is that Eq.\,(\ref{eq:sec2}) usually has \emph{two}
different solutions $\delta E_{n}^{\left(2\right)}$, what means degeneracy
is indeed broken at the second order.

\section{\label{sec:Conclusions-and-Perspectives}Conclusions and Perspectives}

In this work, we gained further insight into the spin noncommutativity
proposed in\,\cite{Gomes:2010xk}. We have shown that the noncommutative
Dirac equation can be derived from an action principle, involving
a Lagrangean which is real and has global phase invariance. This implies,
by Noether's theorem, the existence of a conserved current. The existence
of this current is encouraging because it is important for the physical
interpretation of the model.

We also investigated a very simple quantum mechanical system -- the
Landau problem -- and verified the physical effects of the introduction
of the spin noncommutativity. In this simple setting, it was possible
to derive a Hermitean Hamiltonian from the noncommutative Dirac equation,
which consisted on the standard Dirac Hamiltonian plus a noncommutative
correction of order $\theta$. By using standard perturbation theory,
we shown that there is no correction to the energy levels at first
order in $\theta$. The corrections to the spectrum appear at the
second order in $\theta$, and they break the degeneracy of the excited
states, despite the fact that the noncommutativity does not introduce
further preferred directions in the problem. 

These results are potentially interesting from the phenomenological
point of view. In most treatments of similar problems in noncommutative
quantum mechanics, both in relativistic and non-relativistic regimes,
corrections to the spectra are found already at the first order in
$\theta$\,\cite{nair:2000ii,chaichian:2000si,gamboa:2000yq,Adorno:2009yu},
which can pose very stringent constraints on the noncommutativity
parameters. In our relativistic model, corrections only appear at
order $\theta^{2}$, so the noncommutativity parameters could be less
constrained by existing experimental bounds.

Many questions are still open, however, regarding further developments
in this line of research. Instead of a fixed background field, the
dynamics of the electromagnetic field should be consistently incorporated
in this scenario. More complicated potentials could be investigated,
such as the Coulomb potential, and a particular interesting question
is whether the physical effects of the noncommutativity appear only
at order $\theta^{2}$, as in the Landau problem. Finally, since noncommutativity
is expected to be a very high energy effect, one might investigate
whether a quantum field theory could be defined based on this type
of noncommutativity. The definition of a novel type of noncommutative
quantum field theories, which preserves Lorentz invariance by construction,
would certainly be a very interesting problem.

\vspace{1cm}

\textbf{Acknowledgments.} This work was partially supported by the
Brazilian agencies Conselho Nacional de Desenvolvimento Cient\'{\i}fico
e Tecnol\'{o}gico (CNPq) and Funda\c{c}\~{a}o de Amparo \`{a} Pesquisa
do Estado de S\~{a}o Paulo (FAPESP).

\end{document}